\documentclass[10pt,conference]{IEEEtran}
\ifdefined\XeTeXversion
\usepackage[T1]{fontenc}
\fi
\usepackage{cite}
\usepackage{amsmath,amssymb,amsfonts}
\usepackage{algorithmic}
\usepackage{graphicx}
\usepackage{booktabs}
\usepackage{subcaption}
\usepackage{textcomp}
\usepackage{xcolor}
\usepackage[hyphens]{url}
\usepackage{fancyhdr}
\usepackage{hyperref}
\usepackage[all]{hypcap}
\usepackage[section]{placeins}
\usepackage{tabularx}
\usepackage{array}

\newcommand{\hpcayear}{2027}
\newcommand{\hpcacameraready}{}
\IEEEoverridecommandlockouts
\title{Tetris: Accelerating Composable Test-Time Scaling on GPU-PIM Systems via Adaptive Mapping, Control, and Memory Management}
\newcommand\hpcaauthors{Shengxuan Qiu, Haochen Huang, Tenghao Zhao, and Meng Li
}
\newcommand\hpcaaffiliation{Peking University}
\newcommand\hpcaemail{\{qiusx, hhc, 2501111848\}@stu.pku.edu.cn; meng.li@pku.edu.cn}

\author{
  \ifdefined\hpcacameraready
    \IEEEauthorblockN{\hpcaauthors{}}
      \IEEEauthorblockA{
        \hpcaaffiliation{} \\
        \hpcaemail{}
      }
  \else
    \IEEEauthorblockN{\normalsize{HPCA \hpcayear{} Submission
      \textbf{\#\hpcasubmissionnumber{}}} \\
      \IEEEauthorblockA{
        Confidential Draft \\
        Do NOT Distribute!!
      }
    }
  \fi 
}

\fancypagestyle{camerareadyfirstpage}{%
  \fancyhead{}
  
  \fancyhead[C]{
    \ifdefined\aeopen
    \parbox[][12mm][t]{13.5cm}{\hpcayear{} IEEE International Symposium on High-Performance Computer Architecture (HPCA)}    
    \else
      \ifdefined\aereviewed
      \parbox[][12mm][t]{13.5cm}{\hpcayear{} IEEE International Symposium on High-Performance Computer Architecture (HPCA)}
      \else
      \ifdefined\aereproduced
      \parbox[][12mm][t]{13.5cm}{\hpcayear{} IEEE International Symposium on High-Performance Computer Architecture (HPCA)}
      \else
      \parbox[][0mm][t]{13.5cm}{\hpcayear{} IEEE International Symposium on High-Performance Computer Architecture (HPCA)}
    \fi 
    \fi 
    \fi 
    \ifdefined\aeopen 
      \includegraphics[width=12mm,height=12mm]{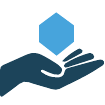}
    \fi 
    \ifdefined\aereviewed
      \includegraphics[width=12mm,height=12mm]{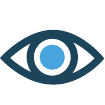}
    \fi 
    \ifdefined\aereproduced
      \includegraphics[width=12mm,height=12mm]{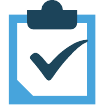}
    \fi
  }
  \fancyfoot[C]{}
}
\begin{document}
\maketitle

\ifdefined\hpcacameraready 
  \thispagestyle{camerareadyfirstpage}
  \pagestyle{empty}
\else
  \thispagestyle{plain}
  \pagestyle{plain}
\fi

\newcommand{\hpcaheight}{0mm}
\ifdefined\eaopen
\renewcommand{\hpcaheight}{12mm}
\fi

\thispagestyle{empty}
\setlength{\textfloatsep}{7.5pt plus 1pt minus 1pt}
\setlength{\dbltextfloatsep}{7.5pt plus 1pt minus 1pt}
\setlength{\floatsep}{6.5pt plus 1pt minus 1pt}
\setlength{\dblfloatsep}{6.5pt plus 1pt minus 1pt}

\begin{abstract}

Test-time scaling (TTS) enhances large language model (LLM) reasoning by spending additional inference-time computation to explore reasoning paths. Specifically, verifier-free TTS leverages LLM-internal signals to guide path expansion, merge, and early termination, and is increasingly adopted in coding and agentic applications. However, composing these actions within a request reshapes its workload and KV-cache state. While hybrid GPU-PIM systems have been proposed to accommodate the heterogeneous compute and bandwidth requirements of verifier-free TTS, prior systems often leverage a static GPU/PIM mapping or coarse-grained remapping, suffering from suboptimal workload split, frequent KV cache allocation/reclamation, and synchronization stalls at reasoning boundaries. 

In this paper, we propose Tetris, an adaptive GPU--PIM architecture with three TTS-aware mechanisms: action-boundary online mapping to rebalance evolving workloads, controller-local KV-state management, and speculative metadata preparation to overlap control coordination with inference. {Across three MoE models and six TTS frameworks, Tetris reaches 8.89$\times$ over optimized GPU execution and averages 2.78$\times$ over AttAcc and 1.87$\times$ over tuned ORCHES at matched action boundaries.}

\end{abstract}

\section{Introduction}
\label{sec:intro}

Test-time scaling (TTS) improves large language model (LLM) reasoning by allocating additional computation during inference~\cite{wei2022chain,welleck2024decoding,zhang2025ttssurvey,snell2025testtime}. It strengthens mathematical and coding reasoning~\cite{wang2024cotdecoding,bon,du2025temperature} and supports multi-step agent workflows~\cite{zhu2025agenttts,hermesagent,claudecode, maservebench}. {OpenAI o1 extends sequential reasoning, while Google Gemini Deep Think explores parallel hypotheses to improve math/coding performance~\cite{openai2024reasoning,google2025deepthink}; Tencent\textquotesingle s Parallel-R1 learns parallel exploration to improve mathematical accuracy~\cite{zheng2026parallelr1}.} TTS makes inference longer and less regular. The cost is acute on edge platforms, where compute throughput, memory bandwidth, capacity, and power are constrained; a single reasoning query can take minutes~\cite{FASTtts}.

\begin{figure}[!t]
    \centering
    \includegraphics[width=\columnwidth]{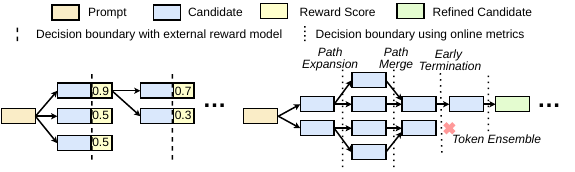}
    \begin{subfigure}[t]{0.46\columnwidth}
        \centering
        \vspace{-14pt}
        \caption{Verifier-based TTS.}
    \end{subfigure}
    \hfill
    \begin{subfigure}[t]{0.52\columnwidth}
        \centering
        \vspace{-14pt}
        \caption{Composable verifier-free TTS.}
    \end{subfigure}
    \vspace{-8pt}
    \caption{High-level comparison of TTS workflows. A reward model scores generated paths in verifier-based execution, whereas online metrics guide path expansion, merge, early termination, and token ensemble without a reward model.}
    \label{fig:intro-workload}
\end{figure}

TTS methods differ in how they decide where to spend this extra computation. Verifier-based methods invoke an external verifier or reward model to score intermediate results, adding model storage, execution, and synchronization overhead~\cite{lightman2024lets,spex,liao2025reward}. \emph{Verifier-free TTS} instead derives decisions from model-internal signals and is attractive for resource-constrained deployment. Recent methods expand reasoning through parallel sampling or verifier-free trees, use consensus or confidence to terminate low-value paths, or ensemble multiple token-level representations within a decoding step~\cite{wang2022self,bon,fu2025deepconf,esc,tuot,threadweaver,roe_hyper_parallel,softthinking,multiplexthinking}. Composable methods further combine several such actions within one request~\cite{qiu2026hyper,adaptivegraph,lbr,xu2025dts}. As illustrated in Figure~\ref{fig:intro-workload}, a request may expand reasoning paths, merge sibling paths derived from the same parent, terminate unpromising paths, and ensemble additional token-level representations within selected decoding steps. We call this regime \emph{composable verifier-free TTS}.

Composable verifier-free TTS is difficult for GPU serving stacks designed around regular autoregressive batches. Action transitions change the number of active paths and the mix of attention over shared prefixes and path-specific KV state. Software runtimes improve paging, prefix reuse, and elastic execution~\cite{pagedattn,radixattn,lmcache,10.1145/3694715.3695948}, but the rapidly changing compute, bandwidth, and state-management demands still expose single-query latency. A PIM-enabled main-memory subsystem is a promising substrate: the GPU handles parallel, compute-intensive operations, while near-bank PIM engines execute bandwidth-bound operations on the same physical memory arrays. AttAcc demonstrates this potential with fixed GPU--PIM operator partitions~\cite{attacc}, while ORCHES adds branch-aware coordination and fragmentation-aware KV reorganization for reasoning workloads~\cite{orches}.


Prior systems are not designed for the coupled workload and state changes introduced by composable verifier-free TTS. AttAcc assumes a largely fixed operator partition between the GPU and PIM, which does not capture how prefix reuse, active-path width, and expert load can change the relative efficiency of the two devices during a request. ORCHES adds adaptive assignment and fragmentation-aware KV reorganization for verifier-guided reasoning, but its coordination focuses on branch-stage workload placement, physical KV layout, and branch pipelining. {Composable verifier-free TTS instead makes device load, logical KV state, and downstream dependencies evolve together within a request. Efficient
support therefore requires these changes to be coordinated jointly, rather than treating placement, compaction, and synchronization as largely separate concerns.}

This execution model exposes three coupled architectural challenges. First, attention and expert placement must track the current compute demand and routed-token distribution; a previously effective GPU--PIM split can become inefficient as these characteristics change. Second, path expansion, merge, and early termination repeatedly change KV ownership and lifetime. Physical compaction can repair fragmented layouts, but logical sharing, survivor publication, and reclamation must also become visible before dependent execution can proceed. Third, the next action and affected paths are resolved online from model signals, so the resulting placement and state changes cannot simply be committed in advance. Hiding this control latency therefore requires preparing state speculatively while preserving a consistent published view.

To address these challenges, we present \textbf{Tetris}, an adaptive GPU--PIM architecture for composable verifier-free TTS. Tetris rebalances computation between GPU and PIM as the reasoning workload evolves, manages changing KV state near memory, and speculatively prepares temporary metadata to overlap control coordination with ongoing inference.

In summary, this work makes the following contributions:
\begin{itemize}
    \item \textbf{Action-boundary GPU--PIM mapping.}
    Tetris rebalances GPU--PIM work from current action descriptors at reasoning-action boundaries.
    {Current prefix reuse and per-expert token loads determine attention splits and expert cuts that balance device completion.}
    \item \textbf{Controller-local KV lifecycle management.}
    Tetris manages evolving path-level KV state near memory.
    {Shared-prefix aliasing, survivor publication, and reclamation update home-controller HATT metadata locally, shortening host-dependent lifecycle transitions.}
    \item \textbf{Decoupled speculative control.}
    Tetris overlaps control preparation with ongoing inference.
    {Predicted indicators drive reversible mapping and KV metadata preparation; the actual action validates local publication before dependent execution proceeds.}

\end{itemize}

\section{Background and Related Work}
\label{sec:back}

\subsection{Verifier-Free TTS Actions and Representative Works}
\label{sec:tts-composable}
\phantomsection\label{sec:moe-tts}
\phantomsection\label{sec:tts-value}
{Verifier-free TTS converts additional inference computation into reasoning gains using model-intrinsic signals. Chain-of-thought prompting elicits intermediate reasoning, improving arithmetic, commonsense, and symbolic problem solving~\cite{wei2022chain}. Self-consistency aggregates diverse paths through agreement, improving GSM8K accuracy by 17.9 percentage points~\cite{wang2022self}. Google\textquotesingle s CoT Decoding ranks reasoning paths by answer confidence~\cite{wang2024cotdecoding}. Self-Certainty uses token distributions for sample selection~\cite{bon}, while entropy-guided temperature selection improves multi-sample inference~\cite{du2025temperature}. Recent methods direct computation through confidence, routing diversity, and path state. DeepConf combines local confidence filtering with weighted voting, reaching 99.9\% AIME25 accuracy with GPT-OSS-120B in its offline 512-trace setting~\cite{fu2025deepconf}. Roster of Experts uses stochastic expert routing and token-level aggregation, improving Mixtral-8×7B's AddSub accuracy from 73.7\% to 79.6\%~\cite{roe_hyper_parallel}. HyPER composes path expansion/reduction, token refinement, and confidence-aware voting, achieving 85.3\% HMMT25 accuracy versus 76.7\% for self-consistency~\cite{qiu2026hyper}. Despite evolving verifier-free TTS policies, recurring path/token actions expose shared execution and KV-lifecycle demands, allowing Tetris to target common primitives rather than any single reasoning algorithm.}

Table~\ref{tab:action-taxonomy} groups execution into three lifecycle-changing \emph{path actions} and token ensemble. \emph{Path expansion} creates concurrent paths from a prompt or active path. \emph{Path merge} selects a sibling using confidence/entropy or combines sibling evidence. \emph{Path early termination} stops low-value or redundant paths. \emph{Token ensemble} aggregates expert outputs, concept representations, or token embeddings within one decoding position. Composable methods combine these actions within one inference flow.

{These actions change active-path populations, shared KV lifetimes, and per-expert token loads. MoE routing supplies token-ensemble alternatives and action-dependent weight reuse. Edge devices must accommodate these changing loads and repeated KV allocation, reclamation, and synchronization within tight compute/memory budgets. This motivates us to study action-driven execution and state management on resource-constrained GPU--PIM systems.}

\begin{table}[!t]
\centering
\fontsize{6.8pt}{7.2pt}\selectfont
\setlength{\tabcolsep}{1.8pt}
\renewcommand{\arraystretch}{0.98}
\caption{Execution-visible actions in representative TTS methods. Checkmarks denote characteristic actions of each method category.}
\label{tab:action-taxonomy}
\begin{tabularx}{\columnwidth}{>{\raggedright\arraybackslash}X *{4}{>{\centering\arraybackslash}p{0.145\columnwidth}}}
\toprule
\textbf{Action} &
\shortstack{\textbf{Path}\\\textbf{expansion}} &
\shortstack{\textbf{Path}\\\textbf{merge}} &
\shortstack{\textbf{Path early}\\\textbf{termination}} &
\shortstack{\textbf{Token}\\\textbf{ensemble}} \\
\midrule
\textbf{Diagram} &
\raisebox{-0.43\height}{\includegraphics[height=0.42in]{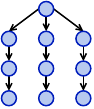}} &
\raisebox{-0.43\height}{\includegraphics[height=0.42in]{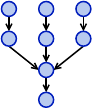}} &
\raisebox{-0.43\height}{\includegraphics[height=0.42in]{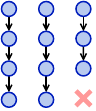}} &
\raisebox{-0.43\height}{\includegraphics[height=0.42in]{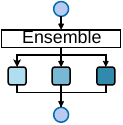}} \\
\textbf{Compute pressure} & High & Low & Low & Extra High \\
\textbf{KV allocation pressure}  & High & Low    & Low  & Low \\
\textbf{KV reclamation pressure} & Low  & Medium & High & Low \\
\midrule
\multicolumn{5}{c}{\textbf{Representative TTS methods}} \\
\midrule
\shortstack[l]{\textit{Parallel reasoning}\\[-1pt]\cite{wang2022self,usc,cisc,bon,nativeparallel}}
& $\checkmark$ & $\times$ & $\times$ & $\times$ \\
\shortstack[l]{\textit{Early-exit reasoning}\\[-1pt]\cite{fu2025deepconf,esc,reasoningaware-sc,relysc,deepprune}}
& $\checkmark$ & $\times$ & $\checkmark$ & $\times$ \\
\shortstack[l]{\textit{Token ensemble}\\[-1pt]\cite{roe_hyper_parallel,softthinking,multiplexthinking}}
& $\times$ & $\times$ & $\times$ & $\checkmark$ \\
\shortstack[l]{\textit{Tree/thread reasoning}\\[-1pt]\cite{yao2023tree,tuot,threadweaver}}
& $\checkmark$ & $\checkmark$ & $\checkmark$ & $\times$ \\
\shortstack[l]{\textit{Composable reasoning}\\[-1pt]\cite{qiu2026hyper,adaptivegraph,prisme,lbr,xu2025dts}}
& $\checkmark$ & $\checkmark$ & $\checkmark$ & $\checkmark$ \\
\bottomrule
\end{tabularx}
\vspace{-2pt}
\end{table}

\subsection{Target GPU--PIM Substrate}
\label{sec:target-substrate}
Processing-in-Memory (PIM) places lightweight compute near memory banks for bandwidth-bound operators~\cite{5,17,34,41}. Tetris targets a package-integrated GPU--HBM3-PIM SoC in which PIM-enabled DRAM serves as GPU main memory. Near-bank GEMV engines operate on the weight and KV arrays also accessed by the GPU. GPU requests traverse an on-package fabric to a Memory Controller Die (MCD), which drives the HBM channels and arbitrates GPU accesses, PIM commands, and background KV movement. The evaluated design combines an Orin-anchored GPU-side service budget with an AttAcc-derived HBM-PIM subsystem; these are separate architectural parameters. Sections~\ref{sec:Tetris-hw-overview} and~\ref{sec:eval-setup} specify the integration contract and bandwidth hierarchy.

\subsection{Related Work}
\label{sec:related-work}

AttAcc~\cite{attacc} targets batched generative inference with a server-class xPU and HBM-PIM accelerator: bank GEMV and buffer-die reduction/softmax execute bandwidth-bound attention, while the xPU executes FC layers. Its fixed operator partition and cross-device pipeline improve batched decoding. ORCHES~\cite{orches} targets edge reasoning on a collaborative GPU--PIM system with a memory controller and near-bank GEMV engines. It adapts workload assignment, pipelines verifier-guided branch rounds, and reorganizes fragmented KV storage.

{Table~\ref{tab:prior-art} distinguishes their coordination semantics. Path actions jointly reshape prefix reuse, expert loads, KV ownership, and downstream dependencies. ORCHES combines branch-round placement, physical compaction, and post-decision control installation; Tetris coordinates action-aware mapping, controller-local logical KV transitions, and reversible preparation using action descriptors and PathIDs. Matching coordination frequency alone therefore leaves placement imbalance and exposed state/control stalls.}

\begin{table}[!t]
\centering
\caption{{Work selection, state management, and next-phase preparation in heterogeneous systems.}}
\label{tab:prior-art}
\begingroup
\tiny
\setlength{\tabcolsep}{1.6pt}
\setlength{\fboxsep}{2pt}
\renewcommand{\arraystretch}{1.08}
{%
\begin{minipage}{\columnwidth}
\begin{tabularx}{\linewidth}{>{\raggedright\arraybackslash}p{0.16\linewidth} *{3}{>{\raggedright\arraybackslash}X}}
\toprule
\textbf{System} & \textbf{Mapping} & \textbf{KV management} & \textbf{Execution overlap} \\
\midrule
AttAcc~\cite{attacc} & Static operator split & Attention-oriented layout & GPU--PIM pipeline \\
ORCHES~\cite{orches} & Branch-level adaptive mapping & Physical compaction & Verifier-guided branch pipeline \\
\textbf{Tetris} & \textbf{Action-level; reuse/load-aware} & \textbf{Controller-local logical updates} & \textbf{Reversible preparation; validated local commit} \\
\bottomrule
\end{tabularx}
\end{minipage}}
\endgroup
\vspace{-2pt}
\end{table}

\phantomsection\label{sec:related-alternatives}
HD-MoE~\cite{11240984} and its extension HDA-MoE~\cite{11690625} optimize MoE inference on 3D near-memory platforms through hybrid tensor/expert placement and dynamic scheduling; HDA-MoE further incorporates hardware-aware routing. These methods address expert deployment and communication on distributed near-memory nodes. Tetris instead coordinates action-dependent attention/expert placement with shared KV lifecycles and control publication on a GPU--HBM-PIM substrate.

{FastTTS~\cite{FASTtts} is a vLLM-based edge serving system combining speculative beam extension, generator/verifier memory allocation, and prefix-aware scheduling. Its beam and generator/verifier policies target software serving on a single GPU; our main comparison fixes verifier-free action traces on common GPU--PIM resources. Our software baselines provide paging/prefix reuse~\cite{pagedattn,radixattn,lmcache}, while ORCHES supplies adaptive placement and branch pipelining as related-mechanism controls.}

\section{Motivation and Bottleneck Analysis}
\label{sec:motivation}

To examine the bottlenecks in Sec.~\ref{sec:intro}, we use SC, DeepConf, RoE, confidence-guided ToT, and HyPER from Table~\ref{tab:action-taxonomy}. {Framework projections retain one representative Qwen3-30B-A3B trace\textquotesingle s context progression while applying each framework\textquotesingle s action/width semantics. HyPER retains the source sequence; other views project the relevant actions from that sequence. The expert diagnostic uses held-out OLMoE routing.}

Placement and KV panels replay the recorded framework traces on modeled AttAcc and ORCHES, with the diagnostic windows and normalization specified below.
\phantomsection\label{sec:workload-dynamics}
Path-count changes are rare but large: 12/450 before/after snapshots across six Qwen3/DeepSeek/OLMoE traces ($W=32/64$) change count, with $|N_{\mathrm{after}}-N_{\mathrm{before}}|/\max(N_{\mathrm{before}},1)=1.0$--2.125. These bursts motivate diagnostic windows; Sec.~\ref{sec:eval-main} evaluates complete queries.

\begin{figure}[!t]
\centering
\includegraphics[width=\columnwidth]{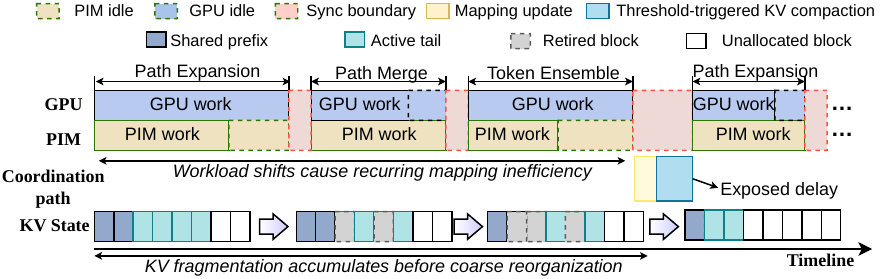}

\caption{ORCHES coordination under composable verifier-free TTS. Actions change placement and KV state between branch-round updates; reactive compaction can expose delay.}
\label{fig:orches-residual}

\end{figure}

The replay exposes action-driven GPU--PIM imbalance, rapid KV-lifecycle changes, and control-boundary synchronization. Figure~\ref{fig:orches-residual} illustrates branch-round assignment and reactive compaction.

\subsection{Action-Driven GPU--PIM Load Imbalance}
\label{sec:motivation-load}
\phantomsection\label{sec:a-figure-clarifications}

\begin{figure}[!t]
\centering
\begin{subfigure}[t]{0.49\columnwidth}
\centering
\includegraphics[width=\linewidth]{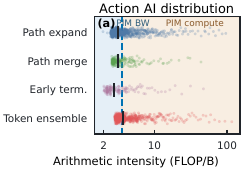}
\phantomsubcaption
\label{fig:dynamic-ai-distribution}
\end{subfigure}\hfill
\begin{subfigure}[t]{0.49\columnwidth}
\centering
\includegraphics[width=\linewidth]{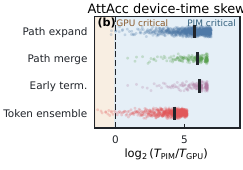}
\phantomsubcaption
\label{fig:dynamic-device-skew}
\end{subfigure}

\includegraphics[width=0.64\columnwidth]{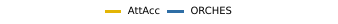}

\begin{subfigure}[t]{0.315\columnwidth}
\centering
\includegraphics[width=\linewidth]{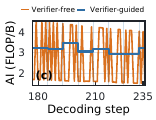}
\end{subfigure}\hfill
\begin{subfigure}[t]{0.315\columnwidth}
\centering
\includegraphics[width=\linewidth]{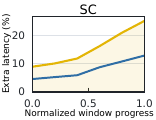}
\end{subfigure}\hfill
\begin{subfigure}[t]{0.315\columnwidth}
\centering
\includegraphics[width=\linewidth]{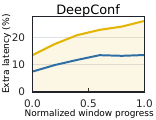}
\end{subfigure}

\begin{subfigure}[t]{0.315\columnwidth}
\centering
\includegraphics[width=\linewidth]{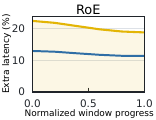}
\end{subfigure}\hfill
\begin{subfigure}[t]{0.315\columnwidth}
\centering
\includegraphics[width=\linewidth]{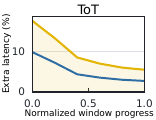}
\end{subfigure}\hfill
\begin{subfigure}[t]{0.315\columnwidth}
\centering
\includegraphics[width=\linewidth]{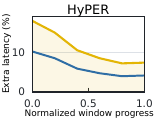}
\end{subfigure}

\caption{Action-level placement mismatch: (a) action arithmetic intensity (Action AI, operations/byte); (b) fixed-partition device-time ratio; (c) action trace and stage projection. {Dots are action-associated steps, jitter separates overlaps, and ticks mark medians. The remaining panels use six-step windows of highest mean ORCHES mismatch; progress spans 0--1 and extra latency is $100(T/T_{\min}-1)$ against the replay's per-step lower envelope.}}
\label{fig:dynamic-placement}

\end{figure}

\begin{figure}[!t]
\centering
\begin{subfigure}[t]{0.49\columnwidth}
\centering
\includegraphics[width=\linewidth]{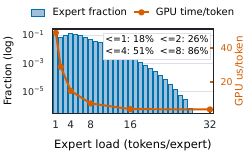}
\par\vspace{-2pt}\captionsetup{skip=0pt}\caption{}
\label{fig:expert-load-distribution}
\end{subfigure}\hfill
\begin{subfigure}[t]{0.49\columnwidth}
\centering
\includegraphics[width=\linewidth]{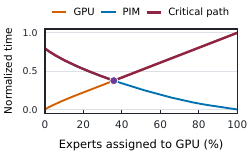}
\par\vspace{-2pt}\captionsetup{skip=0pt}\caption{}
\label{fig:expert-load-split}
\end{subfigure}

\caption{Expert-load diagnostic: (a) invocation fraction and GPU cost; (b) assignment split. {Experts enter the GPU in descending load, times normalize to all-GPU execution, and the marker minimizes device makespan.}}
\label{fig:fanin-motivation}

\end{figure}

{Figure~\ref{fig:dynamic-ai-distribution} shows action-dependent work/traffic distributions, linking changing prefix reuse and expert loads to separate attention and expert placement. Path expansion changes both active width and the shared-prefix/private-tail mix; token ensembles change routed-token counts and weight reuse. These changes shift device completion even when the operator type stays the same. Figure~\ref{fig:expert-load-distribution} shows that most expert invocations have at most eight routed tokens, limiting GPU batching efficiency. Hot experts amortize GPU weight fetch, whereas cold experts benefit from near-bank access. Assigning experts in descending load gives a minimum normalized makespan of 0.376 at 35.8\% GPU assignment (Fig.~\ref{fig:expert-load-split}), motivating T1's expert cut alongside reuse-aware attention mapping. The sweep measures expert-phase completion; its query-time contribution also depends on attention, lifecycle, and control costs.}

Figure~\ref{fig:dynamic-device-skew} uses the same samples: AttAcc assigns attention to PIM and FC/experts to GPU, so positive/negative skew identifies the PIM/GPU tail. \phantomsection\label{sec:device-skew-clarification}{For each sampled step, $T_{\rm PIM}/T_{\rm GPU}$ compares PIM attention with GPU FC/experts. Figure~\ref{fig:main_result} compares complete-query latency against the faster optimized GPU stack executing both partitions. PIM attention can take longer than GPU FC/experts yet less time than GPU attention, allowing positive skew and query speedup together. Peak FLOPS/bandwidth are ceilings; operator work, arithmetic intensity, reuse, bank utilization, and internal versus GPU-visible bandwidth determine achieved service.}

Figure~\ref{fig:dynamic-placement}(c) contrasts action-level intensity with an eight-step projection; the remaining panels quantify high-mismatch windows, showing that the preferred GPU--PIM split depends on current prefix reuse and expert load rather than a fixed or previously installed mapping.

\subsection{Dynamic KV Lifecycles}
\label{sec:motivation-lifecycle}

Action transitions also rapidly reshape KV ownership and lifetime. Path expansion forks shared prefixes into private tails, while path merge and path early termination discard most children. Repeating these transitions interleaves short-lived tails, surviving KV, and free holes, reducing usable capacity and disrupting balanced bank-level PIM access. 

\begin{figure}[!t]
\centering
\begin{subfigure}[t]{0.49\columnwidth}
\centering
\includegraphics[width=\linewidth]{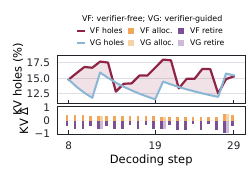}
\par\vspace{-2pt}\captionsetup{skip=0pt}\caption{}
\label{fig:kv-fragmentation}
\end{subfigure}\hfill
\begin{subfigure}[t]{0.49\columnwidth}
\centering
\includegraphics[width=\linewidth]{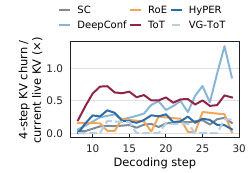}
\par\vspace{-2pt}\captionsetup{skip=0pt}\caption{}
\label{fig:path-state-events}
\end{subfigure}

\begin{subfigure}[t]{0.49\columnwidth}
\centering
\includegraphics[width=\linewidth]{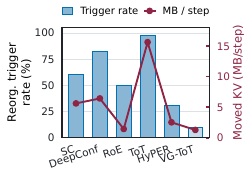}
\par\vspace{-2pt}\captionsetup{skip=0pt}\caption{}
\label{fig:kv-migration}
\end{subfigure}\hfill
\begin{subfigure}[t]{0.49\columnwidth}
\centering
\includegraphics[width=\linewidth]{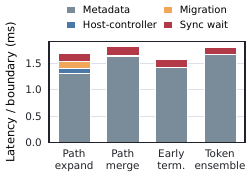}
\par\vspace{-2pt}\captionsetup{skip=0pt}\caption{}
\label{fig:boundary-latency}
\end{subfigure}

\caption{KV-lifecycle diagnostic: (a) holes/events; (b) rolling churn; (c) compaction; (d) boundary cost. {Holes divide by allocated KV; signed events count thousands of blocks against an eight-step guided projection. Churn divides four-step allocated-plus-retired KV by current live KV; compaction reports trigger frequency and MiB/step. Boundary bars average exposed metadata, host-control, migration, and synchronization over selected action windows.}}
\label{fig:kv-control-pressure}

\end{figure}

Path changes require allocation, mapping, and reclamation to become visible before execution continues. The illustrative ORCHES-policy diagnostic in Fig.~\ref{fig:kv-control-pressure} uses $\beta=0.2$ for holes/allocated reasoning memory. Compaction competes with inference for bandwidth and banks.

\phantomsection\label{sec:software-residual}
{Independent of remapping frequency, path-lifecycle changes still require software to publish prefix aliases, survivor descriptors, and reference counts before dependent execution can proceed. The resulting publication/acknowledgment path remains exposed even for zero-byte KV transitions, motivating controller-local handling of logical KV state.}
Figure~\ref{fig:kv-migration} further shows ToT moving about $12\times$ the guided projection's KV, favoring prefix-preserving metadata edits over repeated data reorganization.

\subsection{Control-Boundary Synchronization}
\label{sec:motivation-control}

Verifier-free decisions occur at frequent action boundaries, where the next phase depends on model-internal indicators and the resulting active PathIDs. Mapping installation, path-state publication, lifecycle preparation, and controller metadata updates must therefore wait for that decision, serializing the boundary even for zero-byte KV transitions.

ORCHES pipelines speculative branches across verifier-guided stages. Figure~\ref{fig:boundary-latency} shows path merge's large metadata cost; Secs.~\ref{sec:eval-tech2}--\ref{sec:eval-tech3} distinguish lifecycle service from the subsequent control path.

This motivates preparing tentative placement/path edits during decoding and validating them against the observed action before local publication.

\section{Tetris Design}
\label{sec:Tetris-design}

\begin{figure*}[!t]
    \centering
    \includegraphics[width=0.95\textwidth]{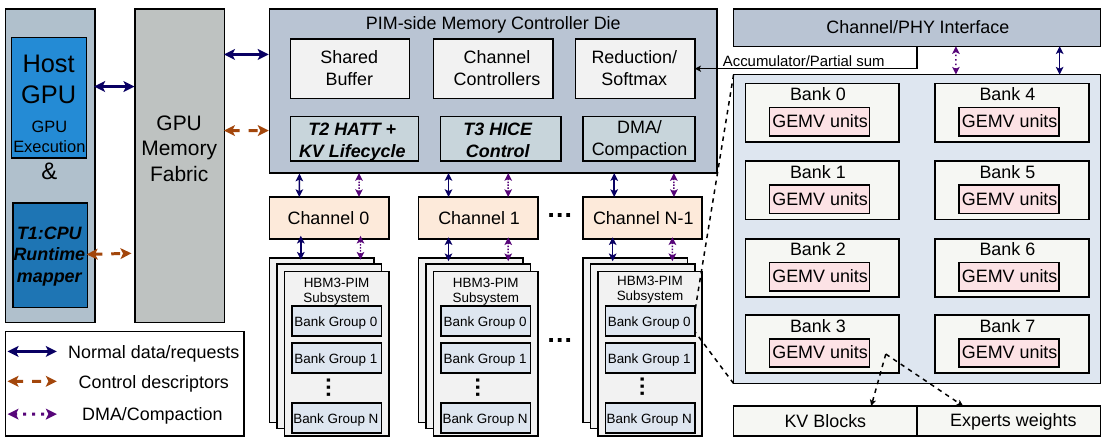}
    \begin{subfigure}[t]{0.62\textwidth}
        \centering
        \caption{Overall GPU--PIM architecture and memory hierarchy.}
        \label{fig:Tetris-hardware-overall}
    \end{subfigure}
    \hfill
    \begin{subfigure}[t]{0.36\textwidth}
        \centering
        \caption{Representative PIM bank group.}
        \label{fig:Tetris-hardware-bank-group}
    \end{subfigure}
    \caption{Proposed Tetris architecture: the GPU--MCD fabric and MCD--HBM channels are distinct interfaces. Bank-local PIM operands remain in memory; GPU requests, exchanged results, and control traffic use the shared hierarchy (Sec.~\ref{sec:Tetris-hw-overview}).}
    \label{fig:Tetris-hardware}
\end{figure*}

\begin{figure}[!t]
    \centering
    \includegraphics[width=\columnwidth]{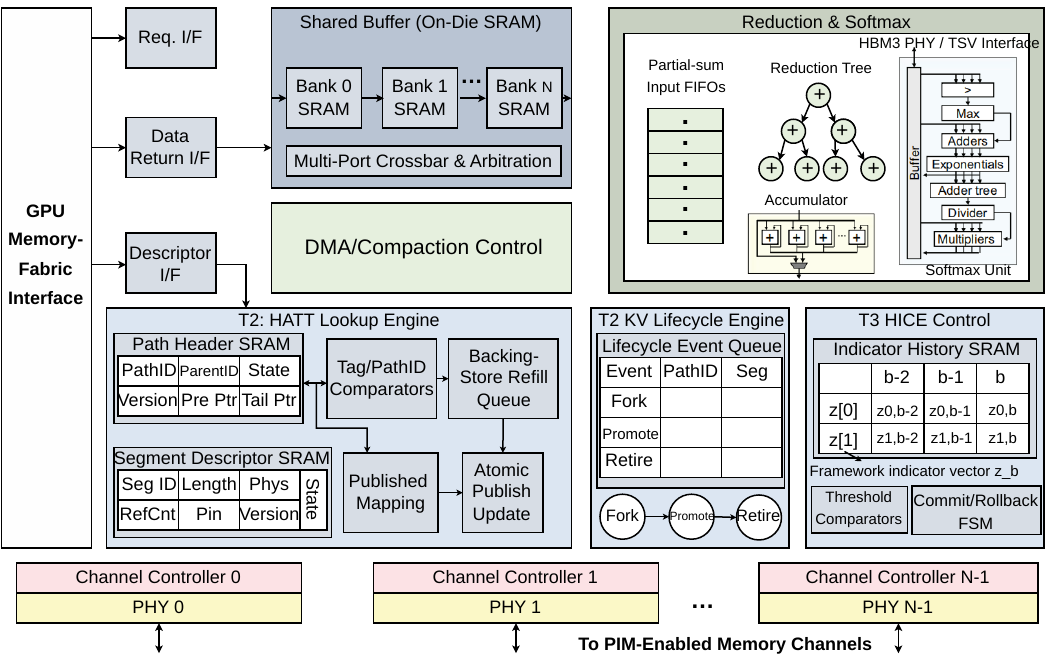}
    \vspace{-6pt}
    \caption{Tetris Memory Controller Die with shared buffering, reduction/softmax, HATT-based translation and KV lifecycle, HICE control, and DMA/compaction.}
    \label{fig:Tetris-controller}
\end{figure}

\subsection{Hardware Architecture Overview}
\label{sec:Tetris-hw-overview}

Figure~\ref{fig:Tetris-hardware} shows the proposed package-integrated GPU--PIM SoC. HBM3-PIM provides one physical storage pool for weights and KV blocks, accessed either by GPU kernels or by near-bank engines. The architecture has two distinct interfaces: a GPU-facing on-package request/data fabric terminating at the MCD, and memory-facing HBM3 channel PHYs connected to the stacked dies through the package and TSV hierarchy. The MCD fans requests across four stacks. Channel control, shared buffering, reduction/softmax, and bank GEMV follow AttAcc~\cite{attacc}; Tetris adds CPU mapping and controller-local lifecycle and speculative control.

\noindent\textbf{GPU--memory interface contract.}
GPU memory requests and CPU-submitted action/PathID descriptors reach the MCD over the SoC fabric. PIM commands are issued by the memory controller, while the bulk weight/KV operands stay at their physical bank locations; activations, partial results, and control records cross the relevant interfaces. A producer must complete before dependent GPU or PIM work consumes its results. We assume the SoC memory system provides the visibility and ordering required at these handoffs; HATT pin/version checks govern KV lifetime and publication. Evaluation represents the interface through its service bandwidth, transferred bytes, arbitration, and exposed synchronization dependencies. The target is a custom integrated SoC using Orin-calibrated GPU rates, with HBM3-PIM replacing conventional main memory; the measured Orin supplies the GPU execution anchor.

\noindent\textbf{Bandwidth hierarchy.}
The GPU-facing fabric has a 204.8-GB/s nominal service budget. The four HBM stacks have a larger aggregate channel-side reference bandwidth, and their bank-local engines expose still greater aggregate operand bandwidth. Memory-side capacity and parallelism therefore need not imply an equally wide GPU-facing interface. GPU reads and returned results remain constrained by the front-end path; PIM-local operand reads bypass that path but compete for banks and command resources. Section~\ref{sec:bandwidth-assumptions} gives the numerical derivation and explains its role in performance comparisons.

\noindent\textbf{Host GPU and CPU runtime.}
The GPU executes dense feed-forward layers, routing, and non-MoE linear work. T1 executes on the CPU, assigning shared-prefix attention and activated experts between GPU and PIM; path-private attention remains near memory. The runtime submits placements and PathID descriptors to the MCD.

\noindent\textbf{Memory Controller Die.}
The MCD resolves logical PathIDs and KV segments to physical locations (Fig.~\ref{fig:Tetris-controller}). HATT manages aliases and canonical reference/pin/version state; HICE prepares tentative descriptors and validates their publication. Shared SRAM buffers operands and partial results for reduction/softmax. Channel arbitration schedules GPU requests, PIM commands, DMA, and metadata accesses; background compaction yields to foreground execution. Path headers, queues, and tentative state reside in controller SRAM, with larger metadata structures in memory-resident backing.

\noindent\textbf{PIM Memory Dies.}
Dies contain channels, bank groups, and banks with fixed-function GEMV engines using weights and KV. Bank-group accumulators combine partial results before controller-side reduction. Expert weights occupy bank groups to reduce token broadcast while exploiting bank parallelism. Section~\ref{sec:eval-setup} gives capacities, rates, timing, and costs.

\subsection{Technique 1: Dynamic Makespan-Balancing Task Mapping}
\label{sec:Tetris-tech1}
T1 maps routable work at each action boundary (Sec.~\ref{sec:motivation-load}).
It maps shared-prefix attention and experts across GPU/PIM to minimize predicted makespan from current work/traffic.

\begin{figure}[!t]
    \centering
    \includegraphics[width=\columnwidth]{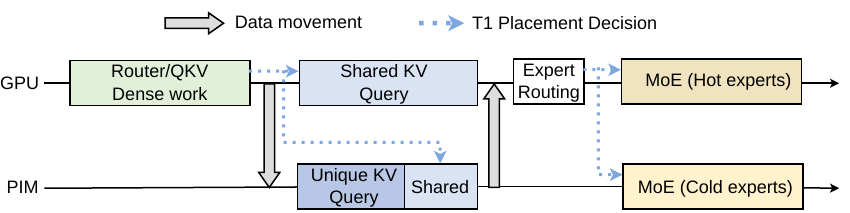}
    \caption{T1 dynamically places shared-prefix attention and activated experts across the GPU and PIM while retaining dense work on the GPU and path-private attention near memory.}
    \label{fig:Tetris-tech1}
\end{figure}
\noindent\textbf{Compact latency abstraction.}
$C$ counts operations (two per MAC) and $M$ transferred bytes at the evaluated precision. For device $d\in\{\mathrm{GPU},\mathrm{PIM}\}$, T1 uses
{\footnotesize
\setlength{\abovedisplayskip}{4pt}
\setlength{\belowdisplayskip}{4pt}
\begin{equation}
\widehat{T}_d(C,M,t)=
\max\!\left(\frac{C}{P_d^{\mathrm{eff}}(t)},\frac{M}{BW_d^{\mathrm{eff}}(t)}\right)+\Delta_d(t).
\label{eq:tech1_latency}
\end{equation}
}
{Here $P_d^{\mathrm{eff}}$ is effective arithmetic throughput in operations/s, $BW_d^{\mathrm{eff}}$ is effective bandwidth in bytes/s, and $\Delta_d$ is exposed scheduling/queueing and memory-state delay in seconds.} GPU rates use the Orin anchors in Sec.~\ref{sec:eval-setup}; PIM rates follow AttAcc/Ramulator2. FP16 activations and INT4 weight bytes enter the descriptors.
\phantomsection\label{sec:platform-parameters}
Platform rates are framework-independent; descriptors supply current work/traffic, prefix reuse, expert loads, and queue/bank state. Retargeting recalibrates device rates (Sec.~\ref{sec:eval-setup}), retaining the action interface/placement rule.

\noindent\textbf{Phase 1: Attention load balancing.}
At boundary $t$, $(C_t^s,M_t^s)$ and $(C_t^u,M_t^u)$ describe shared-prefix and private attention. Split $\alpha$ sends that fraction of shared-prefix work to GPU; PIM executes the remainder and all private attention. Equation~\ref{eq:tech1_latency} predicts loads $L_G^{\mathrm{attn}}(\alpha)$/$L_P^{\mathrm{attn}}(\alpha)$; T1 chooses
{\footnotesize
\setlength{\abovedisplayskip}{4pt}
\setlength{\belowdisplayskip}{4pt}
\begin{equation}
\begin{aligned}
\alpha_t^*
&=\arg\min_{\alpha\in\{0,0.25,0.5,0.75,1\}} \\
&\quad\left[\max\!\left(L_G^{\mathrm{attn}}(\alpha),L_P^{\mathrm{attn}}(\alpha)\right)
+\lambda_{\mathrm{adm}}T_{\mathrm{move}}(\alpha,\alpha_{t-1})\right].
\end{aligned}
\label{eq:tech1_attention}
\end{equation}
}
$T_{\mathrm{move}}$ charges installation/GPU-visible staging at one physical placement. {$\lambda_{\mathrm{adm}}=0.15$ favors stability; installed changes receive full cost. Figure~\ref{fig:parameter-sensitivity}(b) measures the resulting sensitivity.}
\phantomsection\label{sec:attention-grid-sensitivity}
Section~\ref{sec:eval-tech1} evaluates the 5/9/17-point $\alpha$ grids.

\noindent\textbf{Phase 2: Expert-load-aware MoE dispatching.}
For expert $e$, routed-token count $n_e$, per-token work $c$, and weight size $S_e$ give $(C_e,M_e)=(c n_e,S_e)$. Hot experts amortize GPU weight fetch across more tokens; cold experts favor PIM's near-memory execution.

Equation~\ref{eq:tech1_latency} estimates each expert\textquotesingle s GPU/PIM time.

T1 sorts $m$ active experts by descending load. Cut $k$ puts the first $k$ on GPU, the rest on PIM. $L_G^{\mathrm{MoE}}(k)$ includes dense GPU work and those $k$ experts; $L_P^{\mathrm{MoE}}(k)$ includes the remainder. T1 selects
{\footnotesize
\setlength{\abovedisplayskip}{4pt}
\setlength{\belowdisplayskip}{4pt}
\begin{equation}
k_t^*=\arg\min_{0\le k\le m}
\max\!\left(L_G^{\mathrm{MoE}}(k),L_P^{\mathrm{MoE}}(k)\right).
\label{eq:tech1_moe_opt}
\end{equation}
}

Sorting costs $O(m\log m)$; prefix sums evaluate $m+1$ cuts in $O(m)$. One call maps all active paths; Sec.~\ref{sec:eval-tech1} reports costs and transition sensitivity.

\subsection{Technique 2: Controller-Local KV Management}
\label{sec:Tetris-tech2}

PagedAttention/prefix caching support shared non-contiguous KV, but path actions still expose host metadata updates. T2 moves path creation, sharing, selection, and retirement to MCD-local HATT; decoding appends private tails. Arenas/free lists reclaim short-lived allocations; DMA compacts residual live blocks.

\noindent\textbf{Path-aware KV abstraction and HATT.}
Each path has a logical \emph{PathID}. The controller-resident \emph{Hardware Address Translation Table} (HATT) maps it to ordered KV segments backed by physical blocks, caching a path header and segment descriptors.
Headers store PathID, ParentID, State, published Version, and prefix/tail references. Segment descriptors store logical range, physical location, length, and canonical block RefCnt/Pin/version references.

Requests carry PathID/segment offset; one lookup resolves the segment\textquotesingle s commands. Channels look up in parallel; same-entry updates serialize.
Sharing/retirement events update reference counts; ordinary accesses preserve them.
\phantomsection\label{sec:shared-page-protocol}
{Physical-address decoding assigns each shared block one home MCD, the unique owner of its canonical reference/pin/version state; all channel controllers use that serialized state. Forking $p_0$ into $p_1$ aliases block $b$ and raises RefCnt from one to two; readers pin published version $v$ until completion. Tentative aliases stay private, so rollback preserves published mappings/counts; a committed survivor publishes its descriptor atomically. Relocation waits for pins to drain, copies $b$, and publishes its pointer at $v+1$ before freeing the old block. Retiring $p_1$ restores RefCnt to one; retiring $p_0$ permits reclamation after pins drain.}

\begin{figure}[!t]
    \centering
    \includegraphics[width=\columnwidth]{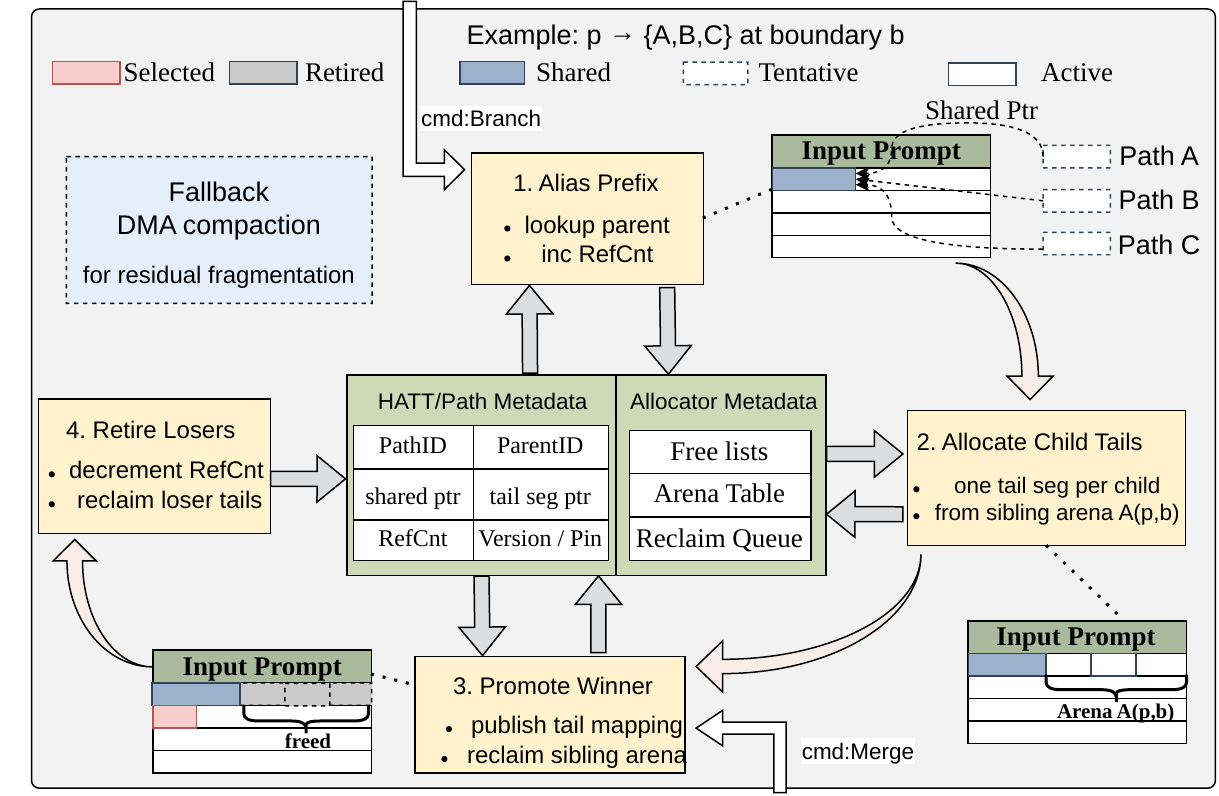}
    \caption{T2 branch-boundary workflow: shared-prefix aliases, sibling-tail allocation, survivor publication, and reclamation; DMA handles residual fragmentation.}
    \label{fig:Tetris-tech2}
\end{figure}

\noindent\textbf{Local lifecycle management.}
T2 handles these events locally. \textbf{Path expansion} aliases the shared prefix, creates a child descriptor, and allocates a new private-tail segment. \textbf{Survivor promotion} publishes the selected tail mapping through controller-local HATT updates. \textbf{Retirement and cleanup} retire obsolete descriptors, reclaim local segments, and compact safe unpinned blocks opportunistically. Figure~\ref{fig:Tetris-tech2} illustrates primitives within a sibling arena, which enables bulk reclamation after survivor publication; nonbranching cases use the reference-count and free-list path.

\noindent\textbf{DMA compaction and visibility-safe remapping.}
DMA compacts unpinned live blocks when fragmentation/churn prevents arena growth; low-priority traffic shares channel/bank bandwidth and defers to foreground kernels.

DMA populates a new location for a valid, unpinned block, atomically publishes its pointer/version in HATT, then reclaims the old location after prior accesses drain. Each GPU/PIM access uses a complete published mapping.

\noindent\textbf{Hardware support.}
T2 comprises HATT, allocation/reclamation, sibling arenas, DMA, and pin/version control.
{HATT caches 256 path headers and sixteen 32-byte descriptors per path: 128~KiB payload plus 16~KiB headers/tags/replacement state. Residency includes active paths and pending siblings across queries; misses or longer chains refill through shared arbitration while lookup remains 5~ns at 200~MHz. Dirty eviction writes published state to backing before slot reuse; pinned/tentative entries defer eviction. Section~\ref{sec:eval-tech2} measures capacity and service costs.}

HATT residency determines locality; physical admission budgets weights, distinct KV, scratch, and backing state.

\subsection{Technique 3: Decoupled Speculative Control}
\label{sec:Tetris-tech3}

At an action boundary, the next phase waits for the actual decision and consistent path, KV, and mapping descriptors. T3 uses the \emph{History-Informed Control Engine} (HICE) to predict next-boundary indicators during the current phase and prepare reversible controller state.

\noindent\textbf{Predictive interface.}
{HICE predicts confidence for DeepConf/ToT; confidence/entropy/consensus/diversity for HyPER; entropy/varentropy for DTS-Greedy branching at its current path budget. DTS completion uses the observed event. Consecutive PathID histories reset on lifecycle changes; insufficient history abstains. Known transitions prepare known signatures; predicted transitions validate signatures at the boundary.}

\noindent\textbf{Conservative indicator prediction.}
Let $\mathbf{z}_b$ denote the normalized indicator vector observed at
action boundary $b$. HICE stores its first- and second-order changes,
\begin{equation}
\Delta_b
=
\mathbf{z}_b-\mathbf{z}_{b-1},
\qquad
\Delta'_b
=
\Delta_b-\Delta_{b-1},
\label{eq:tech3-differences}
\end{equation}
and predicts the next vector using a low-order extrapolator:
\begin{equation}
\hat{\mathbf{z}}_{b+1}
=
\mathbf{z}_b
+
\lambda_b
\left(
w_1\Delta_b+w_2\Delta'_b
\right),
\label{eq:tech3-indicator}
\end{equation}
where $\lambda_b\in\{0,\frac{1}{2},1\}$ sets HICE's extrapolation step.

To detect unstable transitions, HICE measures the average first-order
drift and second-order abruptness:
\begin{equation}
V_b
=
\frac{1}{d}\lVert\Delta_b\rVert_1,
\qquad
A_b
=
\frac{1}{d}\lVert\Delta'_b\rVert_1,
\label{eq:tech3-dynamics}
\end{equation}
where $d$ is the number of indicators exposed by the current
framework. It then selects
\begin{equation}
\lambda_b=
\begin{cases}
1, & A_b\le\tau_A \ \text{and}\ V_b\le\tau_V,\\
\frac{1}{2}, & A_b\le 2\tau_A,\\
0, & \text{otherwise}.
\end{cases}
\label{eq:tech3-gate}
\end{equation}

Low drift/abruptness permits full extrapolation; moderate abruptness contracts the step; $\lambda_b=0$ selects ordinary installation.
{Equal weights $(w_1,w_2)=(0.5,0.5)$ balance drift and curvature with shift-based arithmetic; $(\tau_A,\tau_V)=(0.05,0.01)$ is the global operating point. Figures~\ref{fig:parameter-sensitivity}(d) and~\ref{fig:t3-analysis} evaluate weight/gate tradeoffs and control outcomes.}

\noindent\textbf{Reversible control-state preparation.}
Admitted predictions feed the framework\textquotesingle s registered rule, producing a tentative \emph{control signature}: action, affected PathIDs, and parameters. T3 prepares private child descriptors, aliases, allocation/reclaim entries; only the observed decision authorizes publication, reclamation, and token execution.

\begin{figure}[!t]
    \centering
    \includegraphics[width=\linewidth]{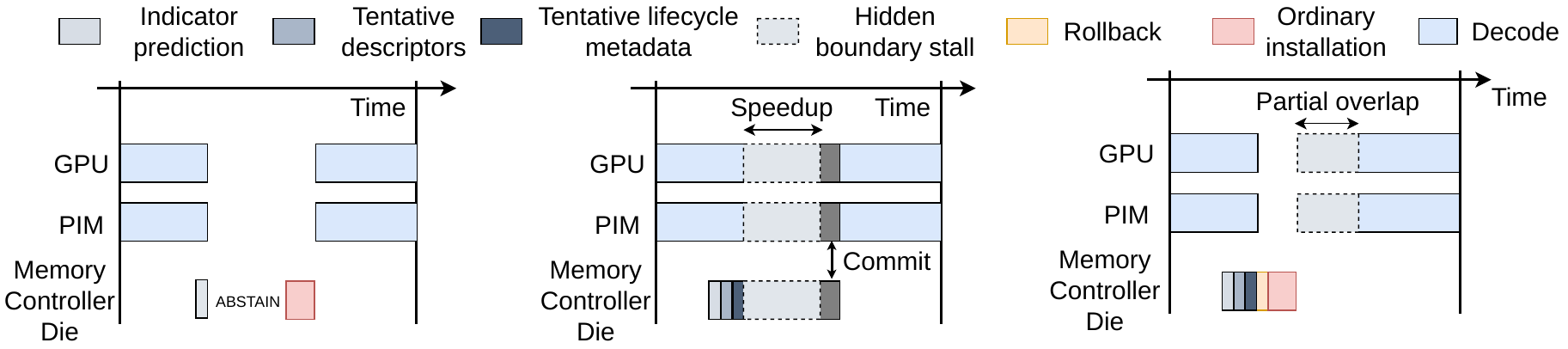}
    \vspace{-5pt}
    \begin{subfigure}[t]{0.32\columnwidth}
        \centering
        \vspace{-5pt}
        \caption{Abstain case.}
        \label{fig:Tetris-tech3-a}
    \end{subfigure}
    \hfill
    \begin{subfigure}[t]{0.32\columnwidth}
        \centering
        \vspace{-5pt}
        \caption{Hit case.}
        \label{fig:Tetris-tech3-b}
    \end{subfigure}
    \hfill
    \begin{subfigure}[t]{0.32\columnwidth}
        \centering
        \vspace{-5pt}
        \caption{Miss case.}
        \label{fig:Tetris-tech3-c}
    \end{subfigure}
    \caption{T3 overlaps reversible metadata preparation with decoding. {Abstention uses ordinary installation; a hit commits locally; a miss rolls back before ordinary installation.}}
    \label{fig:Tetris-tech3}
\end{figure}

\noindent\textbf{Hit, miss, and fallback behavior.}
{At boundary $b+1$, the framework applies its rule to observed indicators; a hit exactly matches the action, affected PathIDs, and parameters, atomically commits descriptors, and acknowledges the GPU. A miss rolls back before ordinary installation; abstention skips preparation and installs after the actual decision (Fig.~\ref{fig:Tetris-tech3}).}

\noindent\textbf{Hardware support.}
HICE comprises an indicator-history buffer, a fixed-point extrapolation datapath, threshold comparators, a framework-rule descriptor store, a
tentative-descriptor queue, and a commit/rollback state machine. It
reuses HATT and lifecycle interfaces for tentative descriptor preparation and atomic commit.

\section{Evaluation}
\label{sec:evaluation}

\subsection{Experimental Setup}
\label{sec:eval-setup}

\noindent\textbf{Hardware configuration.}
Table~\ref{tab:hardware-config} separates GPU-side service, HBM channel resources, and bank-local execution. Four HBM3 stacks provide 4096 banks/64~GiB using AttAcc/Ramulator2's \texttt{HBM3\_8Gb\_2R}/\texttt{HBM3\_5.2Gbps} organization/timing presets~\cite{attacc,ramulator2}. Weights, distinct KV, scratch, and backing state share capacity, with one physical shared-prefix copy. DRAM/buffer areas use their cited processes, separately from the 7-nm controller.

\noindent\textbf{Bandwidth assumptions and their interpretation.}
\phantomsection\label{sec:bandwidth-assumptions}
The 204.8-GB/s GPU-visible budget is anchored to Orin's 256-bit LPDDR5 memory bandwidth~\cite{nvidia_orin_technical_brief}; in the proposed system it parameterizes the GPU--MCD front end. HBM channel capacity is a separate model input: AttAcc's 670.4-GB/s per-stack reference gives $4\times670.4=2681.6$~GB/s, approximately $13.1\times$ the GPU-facing budget. The model intentionally retains the edge-class host service limit while expanding memory-side capacity and bank parallelism. GPU traffic is limited by the slower front end even when the HBM channels have spare capacity.

Bank-local operand bandwidth is a third quantity. AttAcc's power-constrained bank-parallel factor gives $9\times2681.6=24134.4$~GB/s, approximately 24.1~TB/s. Our evaluation retains the configured 24.2-TB/s nominal roof (a 0.27\% difference) for all GPU--PIM systems. This is aggregate near-bank service, rather than bandwidth delivered to the GPU. Activated-bank utilization, DRAM command timing, refresh, and contention determine achieved PIM service. The channel-side reference rate and bank-parallel roof follow AttAcc's performance model; the organization/timing preset specifies the command-level hierarchy and timing. All bandwidths here use decimal bytes/s; capacity uses GiB.

Orin serving measurements use JetPack 6.2.1 (Jetson Linux 36.4.4), MAXN mode, clocks locked with \texttt{jetson\_clocks}, and maximum active cooling. After five warm-up requests, we run three repetitions per question and report median latency from submission to the final token. Cross-request KV state is cleared between trials, retaining intra-query prefix reuse.

\begin{table*}[!t]
\centering
\scriptsize
\setlength{\tabcolsep}{2.5pt}
\caption{Common resources and costs for simulated GPU--PIM systems. The GPU-facing budget, HBM channel reference, and bank-local roof apply to distinct service paths. DRAM area sums distributed dies; buffer compute area is per buffer die (5.6~mm$^2$ across four stacks); added-controller area combines estimated SRAM and placed logic.}
\label{tab:hardware-config}
\begingroup\setlength{\fboxsep}{2pt}
\begin{minipage}{\linewidth}
\centering\begin{tabularx}{\linewidth}{>{\raggedright\arraybackslash}p{0.15\linewidth} >{\raggedright\arraybackslash}X >{\raggedright\arraybackslash}p{0.28\linewidth} >{\raggedright\arraybackslash}p{0.13\linewidth}}
\toprule
\textbf{Block} & \textbf{Configuration} & \textbf{Rate/cost} & \textbf{Source} \\
\midrule
Host GPU & Dense/sparse FP16; shared 64-GiB modeled address space & 43/85~TFLOP/s & Orin-anchored \\
GPU--MCD front end & GPU-visible access and exchanged data; common service budget & 204.8~GB/s & Orin anchor~\cite{nvidia_orin_technical_brief} \\
HBM channel reference & Four stacks; memory-side aggregate, before bank-local scaling & $4\times670.4=2681.6$~GB/s & AttAcc model~\cite{attacc} \\
GPU energy model & Active/resident-idle time; memory bytes & 45/7~W; 28.72~pJ/B & Platform/AttAcc \\
HBM3 organization & 4 stacks $\times$ 16 ch. $\times$ 2 pseudoch. $\times$ 2 ranks $\times$ 4 groups $\times$ 4 banks & 4096 banks; 16~MiB/bank; 64~GiB & AttAcc presets \\
Near-bank execution & 16 MACs/bank; activated-bank accounting & 87.2~TFLOP/s; 24.2~TB/s internal & Common model roofs \\
DRAM timing & All-bank HBM3 refresh; power-constrained MAC spacing & $t_{\mathrm{CK}}\!=\!0.769$~ns; $t_{\mathrm{REFI}}\!=\!3.899$~$\mu$s; $t_{\mathrm{RFC}}\!=\!350$~ns & AttAcc/Ramulator2 \\
DRAM compute area & 4096 GEMV units at 0.094~mm$^2$; 1024 accumulators at 0.036~mm$^2$ & 421.9~mm$^2$ distributed DRAM-die area & AttAcc 1z-nm DRAM \\
Buffer compute area & Softmax (including internal buffers) and accumulation & 1.38 + 0.02~mm$^2$/buffer die & AttAcc 7-nm logic \\
PIM energy model & 0.32~pJ/MAC-equiv.; 4~W active bg.; 3.19~nJ/model cmd. & 13.95~W nom. ALU; activity-based energy & AttAcc-based \\
HATT & $256\times16\times32$~B descriptors plus 16-KiB auxiliary state & 144~KiB SRAM; local lookup/update & Nominal \\
HICE/lifecycle & History, tentative queues, free-list, pin/version state & 64~KiB SRAM; commit/rollback FSM & Nominal \\
Controller total & 208-KiB SRAM plus placed logic; 200~MHz & 0.140~mm$^2$; 0.195~mW logic & SRAM/ASIC \\
SoC/DMA/lifecycle & Transfer/compaction bytes; lifecycle-active time & 10.4/4.4/2~pJ/B; 0.5/5~W Tetris/baseline & Event model \\
Controller events & HATT/HICE/queue/metadata logical events & 8--45~nJ/event; baseline control 0.25~W & Event model \\
DMA/compaction & Descriptor-driven KV movement; foreground-bank avoidance & Low-priority shared-fabric service & Common replay \\
\bottomrule
\end{tabularx}\end{minipage}\endgroup
\end{table*}

\noindent\textbf{Baselines.}
\phantomsection\label{sec:eval-baselines}
{Each configuration normalizes to the faster measured Orin GPU stack, SGLang~\cite{radixattn} or vLLM+LMCache~\cite{pagedattn,lmcache}. AttAcc uses the authors' source; ORCHES's published mapping, pipelining, address caching, and compaction are reproduced on the same backend~\cite{attacc,orches}. All simulated GPU--PIM systems share GPU/PIM compute, the GPU-facing budget, HBM resources, capacity, precision, timing, and traces; the software-only reference is measured on Orin with its native memory system. AttAcc places attention on PIM. ORCHES-A coordinates branch rounds; ORCHES-AB changes only the trigger, retaining placement, compaction, and software KV policies at Tetris's action boundaries. Both ORCHES baselines scan $\beta\in\{0.1,0.3,0.5,0.7,0.9\}$ and use each configuration's minimum-latency setting. Tetris versus AB isolates the combined placement/lifecycle/preparation gains beyond coordination frequency. Figure~\ref{fig:parameter-sensitivity}(a) shows weak query sensitivity because exposed compaction occupies a small phase.}

Tetris uses the global settings in Secs.~\ref{sec:Tetris-tech1}/\ref{sec:Tetris-tech3}; Sec.~\ref{sec:predictor-comparison} specifies query partitioning and frozen calibration scales.

\noindent\textbf{Models, datasets, and workloads.}
\phantomsection\label{sec:eval-workloads}
We evaluate the MoE models Qwen3-30B-A3B~\cite{qwen3technicalreport}, DeepSeek-V2-Lite~\cite{deepseekai2024deepseekv2strongeconomicalefficient}, and OLMoE~\cite{muennighoff2025olmoeopenmixtureofexpertslanguage}, together with the dense Qwen3-8B model~\cite{qwen3technicalreport}. For each Qwen3 model, we evaluate all 30 questions in each of AIME24~\cite{AIME2024}, AIME25~\cite{AIME2025}, and HMMT25~\cite{hmmt_2025}. For DeepSeek and OLMoE, we use one fixed random sample of 50 questions from each of GSM8K~\cite{GSM8K}, MATH500~\cite{hendrycks2021measuringmathematicalproblemsolving}, ARC-Easy, and ARC-Challenge~\cite{ARC}; the same samples are shared by every system.

{MoE uses SC, DeepConf, RoE, confidence-guided ToT, and composable HyPER/DTS-Greedy; dense uses the five path-action frameworks. $W=16/32/64/128$ yields 72 MoE/20 dense configurations. ToT uses confidence-guided BFS with top-$b$ siblings per depth~\cite{yao2023tree,ghasemabadi-etal-2026-guided}. DTS-Greedy~\cite{xu2025dts} composes path expansion, selection, and early termination: it forks top-$k$ paths when next-token entropy $H\leq\theta_H$ and varentropy $V>\theta_V$, freezes branching at the active-path budget, and selects the first EOS/stop completion, retiring other paths.}

Public implementations/reproductions record tokens, expert routes, action boundaries, PathIDs, and KV events at fixed configurations/seeds. Complete-query latencies weight questions equally within each dataset and then dataset means equally.

\noindent\textbf{Precision and partitioned arithmetic.}
\phantomsection\label{sec:eval-numerics}
{Traces and control signals use INT4 weight-only quantization and FP16 activations; replay fixes tokens, routes, and actions across systems. Partitioning preserves attention\textquotesingle s softmax normalization by merging rescaled maxima, masses, and weighted numerators; expert partial sums retain routing/dequantization scales. On identical quantized inputs, 150 FP32 and 150 FP16 accumulation cases have maximum relative L2 errors of $7.82\times10^{-7}$ and 0.554\%, respectively, against FP64.}

{Rounding near a threshold can affect control decisions, so GPU/partitioned pairs recompute confidence, entropy, rankings, and action/PathID/parameter signatures on identical inputs/routes. Seed-0 sampling covers all configurations: 9,200 uniform pairs and 3,060 pairs within 1\% of each threshold's calibration scale. Uniform/near-threshold signature disagreement is 0.076\%/0.098\%.}

{Closed-loop continuations then assess end-to-end quality: GPU/partitioned runs independently update tokens, routes, and actions on 90 Qwen3-30B-A3B prompts from AIME24/25 and HMMT25 under HyPER/ToT, using matched seeds/settings. Answer quality is preserved: accuracy differs by 0 percentage points and mean token cost is 0.998$\times$ GPU.}

\subsection{Simulator Validation}
\phantomsection\label{sec:eval-validation}
Systems replay common operator/KV/boundary events on AttAcc\textquotesingle s backend. GPU, PIM, DMA, and metadata share channel/bank arbitration, charging row conflicts, turnarounds, write drains, and exposed synchronization. Replay charges mapping, backing misses, allocation/reclamation, pins/versions, commit/rollback, DMA, and compaction contention.

{Ramulator2~\cite{attacc,ramulator2} includes all-bank refresh and power-constrained MAC spacing; Fig.~\ref{fig:cost-refresh}(b,c) varies $t_{\mathrm{REFI}}$/spacing at fixed addresses, operation counts, and $t_{\mathrm{RFC}}$.}

\begin{figure}[!t]
\centering
{\includegraphics[width=\linewidth]{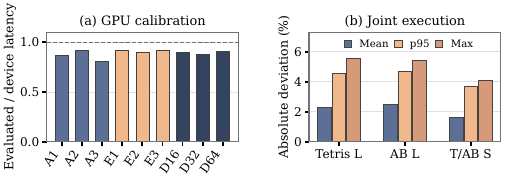}}
\caption{{GPU calibration and joint-execution cross-validation. Panel~(a) reports evaluated/device latency over nine attention, expert, and decode cases; its dashed line marks device timing. Panel~(b) summarizes mean/p95/maximum absolute latency and Tetris/ORCHES-AB speedup deviations over 60 paired configuration means against the detailed contention/synchronization reference.}}
\label{fig:sim-validation}
\end{figure}

GPU calibration (Fig.~\ref{fig:sim-validation}(a)) uses nine matched attention/expert/decode cases, 30 warm-ups, and 100 CUDA-event-timed repetitions with completion synchronization. Mean/max latency deviation is 10.74\%/18.73\%.

{Joint-execution validation (Fig.~\ref{fig:sim-validation}(b)) compares evaluation with an independently constructed detailed simulator scheduling GPU execution, PIM commands, DMA/metadata, arbitration, and synchronization dependencies. Evaluation reconstructs critical paths from calibrated GPU costs and cached, tiled PIM service. Both share recorded operators/actions, resources, and DRAM timing. The 60 configuration means cover six frameworks on Qwen3/DeepSeek at $W=16/32/64/128$ and OLMoE at $W=16/32$. Deviations are $|L_{\rm eval}/L_{\rm ref}-1|$ and $|S_{\rm eval}/S_{\rm ref}-1|$, with $S=L_{\rm AB}/L_{\rm Tetris}$. Paired speedups are less sensitive than absolute timing: mean/max deviations are 1.65\%/4.11\%.}

{Detailed-reference checks add 20 dense and 12 OLMoE $W=64/128$ configurations. Maximum speedup deviation is 2.03\%; the weakest evaluated 1.064$\times$ speedup implies a reference lower bound of 1.04$\times$ over ORCHES-AB.}

\phantomsection\label{sec:command-address-validation}
{Untiled PIM command/address cross-checks give 2.28\% mean/6.79\% max cycle deviation on 18 bank/group/buffer cases ($W=16/32/64$, lengths 128/512). Checks also cover 72 address/broadcast/open-row cases, 18 sustained/refresh cases, and 36 mixed-traffic arbitration tests.}

\phantomsection\label{sec:integrated-timing-envelope}
{Propagating $\pm6.79\%$ PIM-service variation through five HyPER traces (350 boundaries), with GPU timing/actions fixed, changes query latency by at most 5.46\%.}

\subsection{End-to-End Performance}
\label{sec:eval-main}

\begin{figure*}[!t]
\centering
{\includegraphics[width=\linewidth]{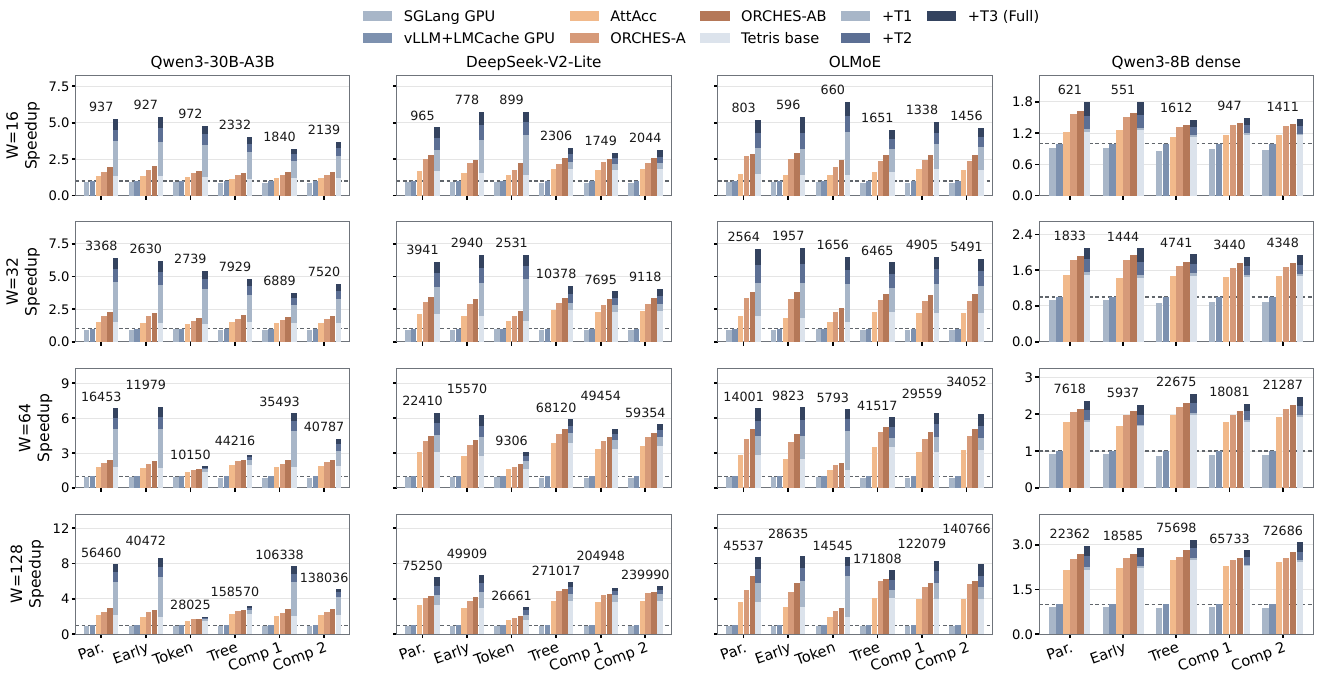}}
\caption{{Single-query speedup at $W=16/32/64/128$: modeled GPU--PIM systems use a common 204.8-GB/s GPU-facing budget and Table~\ref{tab:hardware-config}\textquotesingle s memory-side resources; the denominator is measured Orin software execution. Dense panels use separate scales. Numbers give faster-GPU latency (ms); divide by speedup for system latency. A common GPU denominator makes Tetris/baseline ratios bar-height quotients; headline means are arithmetic means of paired ratios. ORCHES-A/AB use best $\beta$ per point; Tetris stacks add T1--T3. Labels: Par./Early/Token/Tree denote SC/DeepConf/RoE/confidence-guided ToT; Comp~1/2 denote HyPER/DTS-Greedy.}}
\label{fig:main_result}
\end{figure*}

Figure~\ref{fig:main_result} averages paired configuration speedups arithmetically. Against tuned, action-matched ORCHES-AB, Tetris achieves 1.87$\times$/1.10$\times$ on MoE/dense workloads. Matching update frequency removes coordination granularity as a confounding factor, but ORCHES-AB retains its placement, software KV lifecycles, and post-decision control installation. T1 balances prefix reuse and expert loads; T2 localizes logical KV transitions; T3 overlaps reversible preparation. Local KV edits avoid host publication even for zero-byte transitions. Together, they reduce device-completion imbalance and exposed boundary stalls; Sec.~\ref{sec:eval-ablation} separates their contributions.

\begin{figure}[!t]
\centering
{%
\begin{subfigure}[t]{0.485\linewidth}
\centering\includegraphics[width=\linewidth]{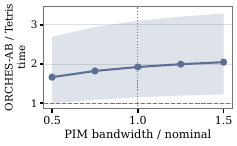}
\caption{{Internal bandwidth.}}
\end{subfigure}\hfill
\begin{subfigure}[t]{0.485\linewidth}
\centering\includegraphics[width=\linewidth]{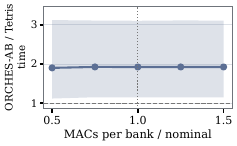}
\caption{{MACs per bank.}}
\end{subfigure}}
\caption{{PIM resource sensitivity over MoE configurations. Solid curves show mean paired ORCHES-AB/Tetris latency ratios; shaded bands show minima and maxima. Bank-local bandwidth (24.2~TB/s nominal) and MACs/bank vary separately with the GPU-facing budget fixed at 204.8~GB/s and GPU compute, capacity, precision, and traces fixed; ORCHES-AB retains its nominal best $\beta$.}}
\label{fig:pim-sensitivity}
\end{figure}

\noindent\textbf{PIM resource sensitivity.}
{At half the nominal PIM bandwidth, the mean ORCHES-AB/Tetris ratio remains 1.66$\times$, although the lower envelope approaches parity (Fig.~\ref{fig:pim-sensitivity}(a)). MAC scaling has a smaller effect: GPU/data-movement dependencies limit critical-path gains, while discrete placements produce small nonmonotonic changes.}

\begin{figure}[!t]
\centering
{%
\begin{minipage}[t]{0.485\linewidth}
\centering\includegraphics[width=\linewidth]{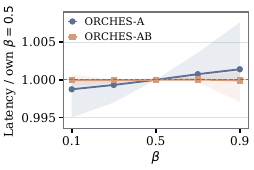}
\par{\scriptsize(a) Compaction.}
\end{minipage}\hfill
\begin{minipage}[t]{0.485\linewidth}
\centering\includegraphics[width=\linewidth]{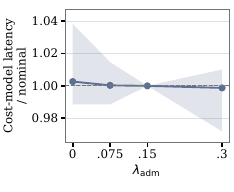}
\par{\scriptsize(b) Admission.}
\end{minipage}
\par\smallskip
\begin{minipage}[t]{0.485\linewidth}
\centering\includegraphics[width=\linewidth]{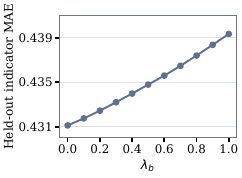}
\par{\scriptsize(c) Fixed extrapolation step.}
\end{minipage}\hfill
\begin{minipage}[t]{0.485\linewidth}
\centering\includegraphics[width=\linewidth]{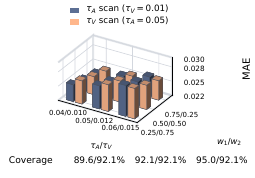}
\par{\scriptsize(d) Weights and gates.}
\end{minipage}}
\caption{{Coordination parameter sensitivity. (a) Mean/min--max latency relative to each configuration\textquotesingle s own $\beta=0.5$. (b) Admission-cost-model latency relative to $\lambda_{\mathrm{adm}}=0.15$, charging full installation. (c) Fixed-step indicator MAE on 84 held-out attempts. (d) MAE (bar tops) and coverage (bottom labels): blue/orange bars scan $\tau_A$/$\tau_V$ with the other gate fixed. All weights/gates share recorded-stream normalization.}}
\label{fig:parameter-sensitivity}
\end{figure}

\noindent\textbf{Interface assumptions and speedup interpretation.}
Figure~\ref{fig:main_result} reports simulated GPU--PIM query latency relative to the faster measured Orin software stack. That comparison combines the benefit of the modeled PIM memory subsystem with Tetris's coordination techniques. The paired ORCHES-AB/Tetris means, 1.87$\times$ on MoE and 1.10$\times$ on dense workloads, instead compare systems with identical GPU-facing and memory-side resources, isolating the coordination benefit at the stated operating point. All GPU-resident operands and exchanged results retain the 204.8-GB/s service constraint; bank-local work uses only the available PIM service.

Increasing the GPU-facing budget can shorten GPU memory service and exchanges, change T1's placement choices, and alter both absolute latency and relative speedup. The 13.1$\times$ channel/front-end ratio and the bank-local bandwidth ratio therefore do not translate directly into query speedups. Figure~\ref{fig:pim-sensitivity} varies PIM-internal bandwidth at a fixed GPU-facing budget; its 1.66$\times$ mean at half nominal PIM bandwidth quantifies that resource sensitivity. A GPU with access to the full aggregate HBM channel rate defines a different operating point requiring recalibrated GPU service and renewed placement/evaluation.

\subsection{Technique Ablation Across MoE and Dense Models}
\label{sec:eval-ablation}

\begin{figure}[!t]
\centering
{\includegraphics[width=\linewidth]{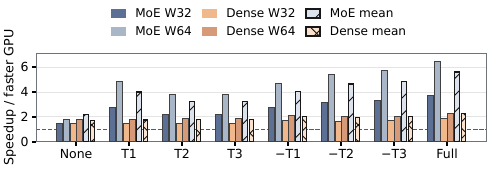}}
\caption{{Standalone and leave-one-out ablations. Four solid bars in each group retain Qwen3 MoE/dense HyPER at $W=32/64$; the two narrow, black-outlined hatched bars give the full MoE ($N=72$) and dense ($N=20$) arithmetic means. All bars use the same per-configuration faster-GPU normalization. T1--T3 enable one technique, $-$T$i$ the other two. The dashed line marks the GPU reference.}}
\label{fig:technique-ablation}
\end{figure}

{Standalone/leave-one-out ablations cover all configurations at fixed resources/traces (Fig.~\ref{fig:technique-ablation}). T$i$'s conditional latency reduction is $1-L_{\mathrm{Full}}/L_{\mathrm{w/o}\,i}$, averaged over paired configurations. The None variant disables T1--T3 on the common substrate. T1-off fixes attention on PIM and FC/experts on GPU; T2-off uses software KV allocation/reclamation/compaction; T3-off installs after the actual decision.}

\noindent\textbf{MoE coordination.} {Figure~\ref{fig:technique-ablation} separates the mechanisms at matched update frequency. T1 alone averages 1.34$\times$ over tuned ORCHES-AB; leave-one-out results attribute added gains to controller-local lifecycle service and speculative preparation beyond finer scheduling.}

\noindent\textbf{Dense-model attribution.} {T2/T3 retain about two-thirds of Full's dense latency saving (Fig.~\ref{fig:technique-ablation}), demonstrating controller-local state/control gains independent of expert routing. Conditional reductions include mechanism interactions.}

{Ablations use fixed controller area; host-side T1 shares the MCD implementation used by T2/T3.}

\subsection{T1: Dynamic Mapping Efficiency}
\label{sec:eval-tech1}

\begin{figure}[!t]
\centering
{\includegraphics[width=\linewidth]{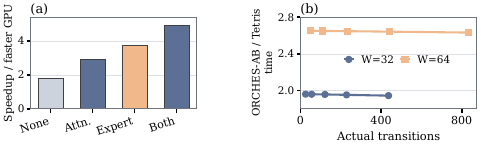}}
\caption{{T1 phase attribution and transition stress. (a) Qwen3/HyPER/$W=64$ with T2/T3 off: No T1 disables both mapping phases; Both enables both. (b) Full Tetris versus ORCHES-AB on Qwen3/HyPER/$W=32/64$, with fixed token work. Transitions count adjacent action-label changes.}}
\label{fig:t1-analysis}
\end{figure}

\noindent\textbf{Attention and expert contributions.}
{With T2/T3 off, neither/attention-only/expert-only/joint mapping achieves 1.79$\times$/2.94$\times$/3.76$\times$/4.90$\times$ over the faster GPU, respectively (Fig.~\ref{fig:t1-analysis}(a)).} Attention uses prefix reuse/width; experts use routed loads. Shared critical paths make gains nonadditive.

{RWMR measures step-wise mismatch; expert makespan (Fig.~\ref{fig:expert-load-split}) weights expert service; query latency (Fig.~\ref{fig:t1-analysis}(a)) also includes attention, fixed work, and boundary dependencies. These metrics weight different execution phases, so expert-level gains need not translate directly into query-level gains.}

\noindent\textbf{Online cost and scale.}
T1 selects on CPU (C++); GPU executes routing, dense, and assigned attention/experts; MCD installs placements and performs T2/T3. WSL/g++-O3 timings use precomputed inputs; the 200-MHz selection reference is separate. Replay charges exposed mapping/installation, lifecycle, and synchronization once.

{Isolated warm-cache CPU microbenchmarks give attention/expert medians of 0.013/0.540~$\mu$s, excluding full-path orchestration, lookup, and submission/installation. Full-path input/descriptor/rate lookup, objective/selection, and submission/installation medians are 1.081/2.892/5.677~$\mu$s. Exposed mapping/installation after overlap is 0.068\% of aggregate query time.}
{Across 2,250 boundaries, selector cost is $\leq$0.0378\% of query time, with 0.81--6.19~$\mu$s medians for 32--256 experts. The 5/9/17-point $\alpha$ grids choose identical five-point minima on twelve saved boundaries.}

\noindent\textbf{Transition frequency.}
{Figure~\ref{fig:t1-analysis}(b) tests Qwen3/HyPER at $W=32/64$ with five pressure levels per width (1/2/4/8/16$\times$). We insert shared-prefix alias fork/prune pairs between original actions, preserving token computation; replay charges their mapping, lifecycle, and control service. Adjacent action-label changes increase from 26/49 to 437/833. At maximum pressure, Full Tetris latency rises 2.27\%/1.93\%, while speedups over ORCHES-AB remain within 1.94--1.96$\times$/2.63--2.65$\times$. Both systems incur extra control work. T1 searches small grids over load summaries; the injected fork/prune pairs edit aliases without moving prefix KV. Their short service relative to decoding keeps speedup ratios stable.}

\subsection{T2: KV Lifecycle Overhead}
\label{sec:eval-tech2}

\begin{figure}[!t]
\centering
\includegraphics[width=\linewidth]{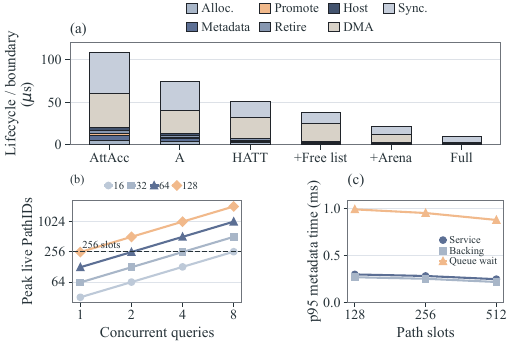}
\caption{T2 lifecycle service and cache capacity. {Panel~(a) averages sampled boundary components; (b) counts active/pending PathIDs versus 256 slots; (c) gives p95 batch service, backing service, and refill-queue wait at $W=64$, eight queries, seed 11. The cache has 16 descriptors/path and 16 refill lanes.}}
\label{fig:t2-analysis}
\end{figure}

Figure~\ref{fig:t2-analysis}(a) decomposes sampled ToT/HyPER lifecycles. HATT converts host-visible mapping/publication into local updates. Free lists and sibling arenas reduce allocation/reclamation; DMA compacts remaining live-block fragmentation. Thus metadata latency falls before compaction service is removed.

\phantomsection\label{sec:boundary-metrics}
{Figures~\ref{fig:boundary-latency}, \ref{fig:t2-analysis}(a), and \ref{fig:t3-analysis}(b) measure selected-window total stalls, sampled KV lifecycle service, and subsequent control preparation/installation, respectively. Their stage averages need not coincide; replay charges each exposed event once on the query dependency path.}

\noindent\textbf{Resident state and backing service.}
{Backing serves HATT overflow (Fig.~\ref{fig:t2-analysis}(b,c)): four $W=64$ queries reach 504 PathIDs beyond 256 slots, while resident lookup remains 5~ns. Eight-query refill-queue-wait p95 is 0.951~ms; long chains refill through backing. Published headers/descriptors restore canonical reference/pin/version state before execution; a shared-page reference checks 1,792 events across three seeds for aliasing, rollback, and reclamation.}

{Despite frequent descriptor refills at W=64/128 (88.8\%/91.9\%; approximately 87,296/184,894 backing operations/query), backing-service p95 remains 0.254/0.537~ms.}

{Within the shared 64-GiB budget, Qwen3-30B-A3B/HyPER/$W=64$ runs with 1/2/4/8 queries peak at 45.67~GiB simultaneous weights, distinct KV, scratch, and backing occupancy, counting shared prefixes once. Capacity-deferred requests wait for admission; all-request wait p95 is 136~ns, including zero-wait requests.}

\begin{figure}[!t]
\centering
{\includegraphics[width=\linewidth]{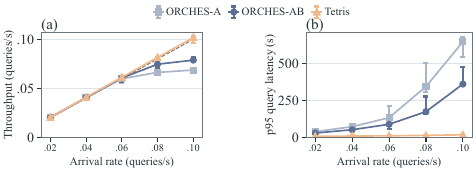}}
\caption{{Open-arrival Qwen3/HyPER/$W=64$ replay. The horizontal axis is injected queries/s. Throughput and query p95 show median/min--max over three seeds and 96 requests/seed; the observation includes drain. The dashed line marks arrival-limited throughput.}}
\label{fig:multi-request}
\end{figure}

\noindent\textbf{Multiple requests.}
\phantomsection\label{sec:eval-multi-request}
{Qwen3/HyPER/$W=64$ queries share HATT, DMA, fabric, and PIM; closed-loop throughput saturates at 2.71$\times$ ORCHES-AB. In open-arrival runs (Fig.~\ref{fig:multi-request}), throughput follows arrivals at low load; near saturation, baseline queueing raises p95 while local lifecycle/preparation sustains more Tetris completions.}

\phantomsection\label{sec:deployment-scope}
{At $W=64$, a 100-request descriptor budget needs 3.125~MiB of payload plus headers, pending siblings, and longer chains. HATT caches resident descriptors; backing and admission handle remaining state in the shared 64-GiB pool.}

\subsection{T3: Synchronization-Wait Hiding}
\label{sec:eval-tech3}

\begin{figure}[!t]
\centering
{\includegraphics[width=\linewidth]{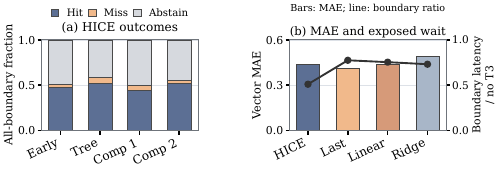}}
\caption{{T3 outcomes and predictor control benefit. Qwen3 at $W=16/32$ uses DeepConf, ToT, HyPER, and DTS-Greedy. Panel~(a) combines widths for all-boundary hit/miss/abstain fractions; (b) averages vector MAE and boundary latency relative to no-T3 installation over eight paired configurations. Last is persistence; all share targets, histories, gates, and attempts.}}
\label{fig:t3-analysis}
\end{figure}

Figure~\ref{fig:t3-analysis}(a) reports hit/miss/abstention over all boundaries.
{The framework-specific inputs in Sec.~\ref{sec:Tetris-tech3} feed registered rules that convert predictions into action signatures, checked against observed actions. SC/fixed RoE use known-state preparation; history-limited or unstable paths use ordinary installation and remain in the abstention count.}

Coverage is $1-P_{\mathrm{abstain}}$ over eligible boundaries; hit/miss rates condition on attempts. Figure~\ref{fig:t3-analysis}(b) computes exposed latency using unconditional probabilities:
\begin{equation}
\mathbb{E}[L]=P_{\mathrm{hit}}L_{\mathrm{hit}}+P_{\mathrm{miss}}L_{\mathrm{miss}}+P_{\mathrm{abstain}}L_{\mathrm{ordinary}}.
\label{eq:t3-exposed-latency}
\end{equation}
Hits charge prediction/exposed preparation/commit; misses add rollback/ordinary installation; abstention charges gating/ordinary installation. The no-T3 reference uses ordinary installation; Eq.~\ref{eq:t3-exposed-latency} counts exposed service.

\phantomsection\label{sec:predictor-comparison}
{HICE, persistence, linear extrapolation, and Ridge share targets, histories, frozen scaling, and gate-defined attempts. Pooled AIME24/25 queries split 30/30 for calibration/test; all paths/frameworks/widths per query stay together. Test weights/gates remain frozen; histories update online. Ridge uses values/first differences with regularization 0.01. Exact action/PathID/parameter matches define hits. Figure~\ref{fig:t3-analysis} shows 30.1--33.9\% lower HICE control latency than each of the three comparators. Persistence's lower MAE shows that action/signature utility, rather than indicator error alone, governs control benefit.}

{Figure~\ref{fig:parameter-sensitivity}(c) scans $\lambda_b$ on 84 chronological held-out attempts with calibration-prefix scales; (d) uses recorded-stream min--max normalization for MAE/coverage over 442 boundaries. $\tau_A$ trades coverage/error; $\tau_V$ selects full/half steps. Gating precedes extrapolation, making coverage independent of $w_1/w_2$.}

{Every tested one-factor setting reduces mean control latency: the highest mean per-configuration control/ordinary-installation ratio (Eq.~\ref{eq:t3-exposed-latency}) is 0.513, with all-boundary hit/miss/abstain fractions 0.467/0.068/0.465. Fig.~\ref{fig:t3-analysis}'s held-out queries/frozen scales use $w_1=0.25/0.5/0.75$ ($w_2=1-w_1$), $\tau_A$/$\tau_V$ at $0.5/1/2\times$ default, and admitted $\lambda_b=0.5/1$.}

\subsection{Power and Energy}
\label{sec:eval-power}

\begin{figure}[!t]
\centering
{\includegraphics[width=\linewidth]{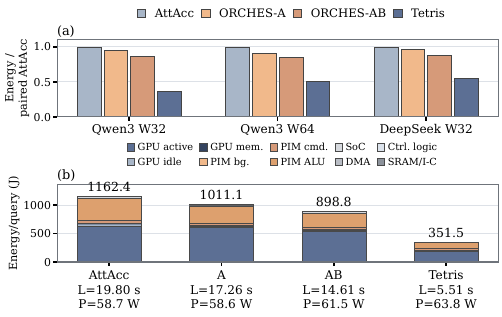}}
\caption{{Paired energy and latency. Panel~(a) averages configuration energy relative to AttAcc over five frameworks per labeled model/width group. Panel~(b) pairs Qwen3/HyPER/$W=64$ component joules with query seconds and average watts on Fig.~\ref{fig:main_result}'s interval.}}
\label{fig:power-energy}
\end{figure}

Query energy sums GPU, PIM/DRAM, SoC, DMA, and controller joules from execution/residency, memory commands, transferred bytes, and metadata events. Shared unit coefficients follow Table~\ref{tab:hardware-config}; controller energy combines post-route active logic with SRAM/interconnect accesses.

{Mean energy is 45.0\% below ORCHES-AB over Fig.~\ref{fig:power-energy}\textquotesingle s 15 paired configurations.} Shorter residency reduces GPU/PIM energy; local lifecycles reduce DMA/control activity.

\phantomsection\label{sec:energy-consistency}
{Energy and latency share traces, submission-to-final-token intervals, and question/dataset weights. Component sums give $E$ and $\bar P=E/L$, satisfying $E_T/E_B=(L_T/L_B)(\bar P_T/\bar P_B)$ per configuration before averaging. Figure~\ref{fig:power-energy}(b) shows lower energy from shorter residency despite higher mean power.}

\begin{figure}[!t]
\centering
{\includegraphics[width=\linewidth]{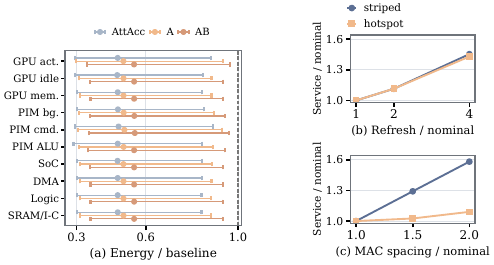}}
\caption{{Energy-cost and sustained-service sensitivity. (a) Means and min/max energy ratios over the 15 paired configurations in Fig.~\ref{fig:power-energy}, varying one shared component coefficient at a time over 0.5--2$\times$ with activity fixed. (b,c) Fixed 396,800-request, 1024-bank HBM3 streams report service time relative to their own nominal timing.}}
\label{fig:cost-refresh}
\end{figure}

\noindent\textbf{Cost and sustained-service sensitivity.}
{Shared-coefficient scans at fixed ledger activity preserve energy ordering. Extra Tetris-only power breaks even at $(E_B-E_T)/L_T$, with minimum per-configuration budget 1.73~W.}

\phantomsection\label{sec:thermal-service}
{In Fig.~\ref{fig:cost-refresh}(b,c), quartering $t_{\mathrm{REFI}}$ and doubling issue spacing yields 2.29$\times$/1.57$\times$ striped/hotspot service time; hotspot contention reduces spacing sensitivity. These sweeps characterize sustained service; package-level thermal transients are not modeled explicitly. Composite SRAM/interconnect/queue events use 8--45-nJ coefficients, with post-route logic power charged separately.}

\subsection{Area and ASIC Implementation}
\label{sec:eval_hw_cost}

\begin{figure}[!t]
\centering
\includegraphics[width=\linewidth]{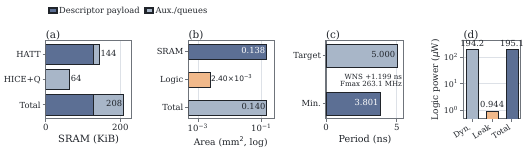}
\caption{Controller footprint and ASAP7 results: (a) SRAM; (b) area; (c) timing; (d) logic power. Area combines estimated SRAM with placed logic; power uses gate-level replay activity.}
\label{fig:area-breakdown}
\label{fig:controller-cost}
\end{figure}

{The nominal controller occupies at most 9.2\% of the combined area with AttAcc's 1.4-mm$^2$ softmax/accumulator per buffer die.} The 208-KiB controller combines 144-KiB HATT and 64-KiB HICE/lifecycle SRAM. A 0.027-$\mu\mathrm{m}^2$/bit 7-nm 6T bitcell~\cite{wu2016_7nm_sram} with $3\times$ peripherals gives 0.138-mm$^2$ SRAM; OpenROAD's 2,396-$\mu\mathrm{m}^2$ routed core gives 0.140-mm$^2$ total (Fig.~\ref{fig:area-breakdown}(a,b)).
{Peripheral factors 2--5 give 0.094--0.232~mm$^2$ total area. Table~\ref{tab:hardware-config} lists inherited DRAM/buffer costs~\cite{attacc}.}

RTL covers lifecycle/HATT, arenas/free lists, HICE/queues, commit/rollback, and DMA. ORFS/OpenROAD 26Q3 uses ASAP7 TT (0.70~V)~\cite{clark2016_asap7}, a 5-ns target, and external SRAM. Figure~\ref{fig:controller-cost}(c,d) reports timing and post-route logic power; final timing, electrical, routing, and antenna checks pass.

Gate-level replay of 100 Qwen3/HyPER/$W=64$ boundaries validates commit/rollback/abstention and DMA completions. VCD activity annotates all post-route pins; extracted-parasitic logic power is 0.195~mW at 200~MHz. Section~\ref{sec:eval-power} adds SRAM/interconnect energy.

\section{Conclusion}
\label{sec:conclusion}

We presented \textbf{Tetris}, a GPU--PIM architecture that jointly coordinates dynamic computation placement, logical KV state, and control dependencies for composable verifier-free TTS. Across MoE workloads, Tetris averages 2.78$\times$ over AttAcc and 1.87$\times$ over tuned, action-matched ORCHES-AB, demonstrating gains beyond coordination frequency while reaching 8.89$\times$ over the faster optimized GPU baseline. Dense-model ablations further show that its benefits extend beyond MoE-specific expert dispatch, while resource, concurrency, and implementation studies characterize robustness and hardware cost. As TTS expands into multi-step reasoning and agent workflows~\cite{zhu2025agenttts,openclaw,hermesagent,claudecode,maservebench}, Tetris provides a common hardware/software interface for coordinating evolving path, KV, and execution state.
\section*{Acknowledgment}
This work was supported in part by the National Key Research and Development Program under Grant 2024YFB4505004, in part by NSFC under Grants 62495102 and 92464104, and in part by the 111 Project under Grant B18001.
\FloatBarrier
\nocite{*}
\bibliographystyle{IEEEtranS}
\bibliography{refs}
\appendices
\section{AI use}

We used OpenAI ChatGPT to assist with English
language polishing, improving clarity and conciseness,
and drafting or refining selected passages in the
manuscript and revision cover letter. The architectural design, technical contributions, implementation, experiments, and reported results were developed by the authors. All AI-assisted content was reviewed and verified by the authors, who take full responsibility for the manuscript.

\end{document}